\documentclass[pra,twocolumn]{revtex4-2}

\usepackage{epsfig,amsmath}
\usepackage{subfigure}
\usepackage{graphicx}
\usepackage{dcolumn}
\usepackage{stmaryrd}
\usepackage{mathrsfs}
\usepackage{pifont}
\usepackage{amsthm}
\usepackage{amssymb}
\usepackage{bm}
\usepackage{latexsym}
\usepackage{color}
\usepackage{hyperref}
\usepackage{multirow}
\usepackage{color}
\usepackage{babel}

\begin{document}
	
\title{Approximate quantum gates compiling with self-navigation algorithm}
\author{Run-Hong He$^{1}$, Ren-Feng Hua$^{2}$, Arapat Ablimit$^{1}$ and Zhao-Ming Wang$^{1}$}
\email{wangzhaoming@ouc.edu.cn}

\address {$^{1}$College of Physics and Optoelectronic Engineering, Ocean University of China, Qingdao 266100, People’s Republic of China\\
$^{2}$ School of Information and Control Engineering, Qingdao University of Technology, Qingdao 266520, People’s Republic of China}

\date{\today}

\begin{abstract}
The compiling of quantum gates is crucial for the successful quantum algorithm implementations. The environmental noise as well as the bandwidth of control pulses pose a challenge to precise
and fast qubit control, especially in a weakly anharmonic system.
In this work, we propose an algorithm to approximately compile single-qubit
gates with arbitrary accuracy. Evaluation results show that the overall
rotation distance generated by our algorithm is significantly shorter than the commonly used $U3$ gate, then the gate time can be effectively
shortened. The requisite number of pulses
and the runtime of scheme design scale up as $\mathcal{O}[\mathrm{Log}(1/\epsilon)]$
with very small prefactors, indicating low overhead costs. Moreover,
we explore the trade-off between effectiveness and cost, and find
a balance point. In short, our work opens a new avenue for efficient quantum algorithm implementations with contemporary quantum technology. 
\end{abstract}
\maketitle

\section{Introduction}
\label{Sec.1}

Owing to the intrinsic properties afforded by quantum mechanics, the
quantum algorithms permit superpolynomial or even exponential speedups
relative to their classical counterparts in solving some important
problems as the input size scales up \cite{qc_nielsen,qc_bb,quantum_supremacy}.
The global race to build quantum computing prototypes is in full swing
now, which is evident by several impressive demonstrations of this
quantum computational advantage successively during the past
few years \cite{google_2019,zuchongzhi1,jiuzhang1,zuchongzhi2,jiuzhang2,zuchongzhi21}.
The full power of quantum computation is supposed to be based on the
universal quantum computer with large scale and error corrected logical
qubits, such as the surface code \cite{surface_code_1}. 

Nevertheless, before the extravagant hardware resources for quantum
error corrected technology within reach, it would take years or even
decades in the noisy intermediate-scale quantum (NISQ) era \cite{NISQ},
in which the actual qubits are not immune to noises and the size of
quantum processor could be faithfully controlled is also relatively
limited. The various sources of noise will impose deleterious impact to the delicate
qubits, then the quantum device fails to produce results with
sufficient fidelity \cite{sc_review_1}. Normally the effects of these noises become worse over time \cite{circuit_approximation}. 

To achieve an improved performance in NISQ device, except for the
effort to eliminate the sources of noise and the endeavor to advance
qubit with reduced noise susceptibility, the optimal
control search  is also an active topic \cite{he_RG,he_DRL}. In particular,
the emergence of circuit-approximation schemes has attracted a lot of
attention. These schemes do not aim to faithfully execute a given algorithm
circuit, but explore an approximated version with fewer gates
and shorter circuit depth, such as in Refs. \cite{circuit_approximation,q_factor}.
These works prove that the approximate circuits could be possible
to outperform the theoretically ideal but deeper circuits 
on contemporary devices. 

However, with too much attention being paid to the circuits optimization,
the approximation to arbitrary unitary gate has been neglected, which
may bring new surprises for further improvement. This provide a promising
approach for exploiting the potential of NISQ devices, especially
in the control over qubit made from a weakly anharmonic oscillator,
where the unwanted leakage transitions to higher states pose another
threat to the implementation of fast and precise quantum gate. For example, in
transmon-type qubit carried by superconducting circuits, spectroscopically,
the non-zero overlaps between the drive and the leakage transition
frequencies arising from finite pulse duration will drive the qubit
out of the Hilbert subspace spanned by the $|0\rangle$ and $|1\rangle$. The
scheme of derivative reduction by adiabatic gate (DRAG) \cite{DRAG_2,DRAG}
is proposed as a typically approach to reduce this population leakage
by modifying the quadrature amplitudes of the microwave drive. While,
in practice, the crucial optimal value of the scaling parameter $\lambda$
in DRAG is sensitive to the pulse distortions and has to be identified
and calibrated repeatedly in the experiments \cite{DRAG_1,DRAG_3}. 

Considering the spectral content of the leakage frequency decays exponentially
with respect to the pulse duration, a naturally alternative to the DRAG for the prevention of
the leakage is using a slower operation with more gentle control pulses,
resulting in the control bandwidth being much less than the anharmonicity. 
Generally speaking, this strategy will extend gate execution time and hence
limit the reliable circuit depth due to the given decoherence time.
While, if the unitary transform is compiled cute enough, it is possible to reduce the overall gate time by finding a short path to compensate the extended time spent in unit rotation distance. In this paper, we provide a feasible route, namely the self-navigation ($SN$) algorithm, to access the approximated
gate compiling. We show that any accuracy can be obtained relative
to the desired precise gate, and the overall distance of the rotation
route is shorter than the commonly used today, such
as the $U3$ gate \cite{virtual_Z_gate,HUAWEI_U3,IBM_U3}.  The $SN$ algorithm delivers the number of elementary-rotations
and the run time both in $\mathcal{O}[\mathrm{Log}(1/\epsilon)]$. This is a
significant enhancement compared to the polylogarithmic in Solovay-Kitaev
algorithm \cite{SK_algorithm}, which outputs a sequence of $\mathcal{O}[\mathrm{Log}^{3.97}(1/\epsilon)]$ elementary-rotations and runs
in $\mathcal{O}[\mathrm{Log}^{2.71}(1/\epsilon)]$ time with the specified accuracy $\epsilon$ to the target gate.

\section{Model}
\label{Sec.2}

The transmon qubit \cite{transmon} is one of the most common qubit
modalities formed by weakly anharmonic oscillators, whose effective
Hamiltonian in the rotating frame reads \cite{sc_review_1,sc_review_2}
\begin{equation}
H_{q}=-\frac{\hbar}{2}\Delta\sigma_{z},\label{eq:1}
\end{equation}
where $\Delta$ denotes the qubit detuning from the frame frequency.
When the qubit resonates with the frame, i.e., $\Delta=0$, the rotation
around the axis in the $XY$-plane can be realized by adding
microwave drive to the qubit for a certain amount of time. The corresponding Hamiltonian
under the rotating wave approximation reads
\begin{equation}
H_{d}=\frac{\hbar}{2}A(\cos\phi\sigma_{x}+\sin\phi\sigma_{y}),\label{eq:2}
\end{equation}
where $A$ refers to the drive amplitude, and $\sigma_{i}$ $(i=x,y,z)$ is the Pauli matrice. The phase of the microwave $\phi$ determines
the rotation axis. For simplicity, we set $\hbar=1$ and take $1/\hbar$
as the time-scale throughout. 

The shift of the qubit frequency $\Delta$ in Eq.~(\ref{eq:1}) leads
to a rotation rate around the $z$-axis and can be tailored by modulating
the flux bias of the SQUID (superconducting quantum interference device)
\cite{Introduction_to_sc}. However, this is not necessary in the experiments
and can be accessed alternatively by leveraging the so-called
``virtual $Z$ gate'' technique \cite{virtual_Z_gate}. This technique can be done by simply
applying a specific phase offset $\phi$ to the microwave signals
for subsequent rotation about axis in the $X$-$Y$ plane. For example,
\begin{equation}
\exp(-i\frac{\theta}{2}[\cos\phi\sigma_{x}+\sin\phi\sigma_{y}])=Z_{-\phi}X_{\theta}Z_{\phi},\label{eq:3}
\end{equation}
\begin{equation}
\exp(-i\frac{\theta}{2}[\cos(\frac{\pi}{2}+\phi)\sigma_{x}+\sin(\frac{\pi}{2}+\phi)\sigma_{y}])=Z_{-\phi}Y_{\theta}Z_{\phi}.\label{eq:4}
\end{equation}
Note that they leave an extra $Z_{-\phi}$ which does not change the outcome
of measurement along $Z$. The implemented virtual $Z$ gate is ``perfect'',
because it requires no additional control pulse and therefore taking
zero-time and having unity gate fidelity nominally. By leveraging
the virtual $Z$ gate technology, one could effectively reduce the
number of overall pulses for the implementation of a quantum gate.

A generic single-qubit gate (ignoring the overall phase) can be realized
by three successive rotations around the $x$- and $z$-axes. These are both native operation in superconducting circuits model \cite{virtual_Z_gate,beginner,IBM_U3}
\begin{equation}
U(\theta,\phi,\lambda)=Z_{\phi}X_{\theta}Z_{\lambda}=\left[\begin{array}{cc}
\cos(\theta/2) & -\mathrm{e}^{i\lambda}\sin(\theta/2)\\
\mathrm{e}^{i\phi}\sin(\theta/2) & \mathrm{e}^{i(\lambda+\phi)}\cos(\theta/2)
\end{array}\right],\label{eq:5}
\end{equation}
where $\theta$, $\phi$ and $\lambda$ represent $3$ Euler angles.
Combined with the identity,
\begin{equation}
X_{\theta}=Z_{-\pi/2}X_{\pi/2}Z_{\pi-\theta}X_{\pi/2}Z_{-\pi/2},\label{eq:6}
\end{equation}
Eq.~(\ref{eq:5}) can be reexpressed as 
\begin{equation}
U(\theta,\phi,\lambda)=Z_{\phi-\pi/2}X_{\pi/2}Z_{\pi-\theta}X_{\pi/2}Z_{\lambda-\pi/2}.\label{eq:7}
\end{equation}
The above is the so-called $U3$ gate technology, a commonly used strategy to compile single-qubit logical operations in the experiments \cite{IBM_U3,HUAWEI_U3}.  As mentioned above, the rotations around $z$-axis
can be included into the microwaves used for the $X_{\pi/2}$ as an
additional phase. That is to say, the requisite total time is always that
used to implement two $X_{\pi/2}$ pulse. Can we find a short rotation path, thereby reduce the gate time and suppress the debilitating decoherence
effects over time? In this paper, by using the $SN$ algorithm a shorter rotation distance has been obtained compared to the $U3$ gate.

\section{The algorithm }
\label{Sec.3}

Arbitrary single-qubit gate can be represented by a rotation around
a particular axis $\overrightarrow{n}$ with a certain angle $\theta$,
\begin{equation}
R_{\overrightarrow{n}}(\theta)=\mathrm{cos}(\frac{\theta}{2})-i\mathrm{sin}(\frac{\theta}{2})(n_{x}\sigma_{x}+n_{y}\sigma_{y}+n_{z}\sigma_{z}).\label{eq:8}
\end{equation}
We firstly consider the rotations around a fixed axis
$\overrightarrow{n}$. Assume the target
rotation angle is $\theta_{T}$ and the current angle is $\theta_{t}$.
Here the pulse amplitude and duration are included  into the rotation
angle. To determine the appropriate next rotation angle, we define
the fidelity $F$ between the two unitaries using the Hilbert-Schmidt
distance \cite{distance_measures,DRL_for_CNOT}
\begin{equation}
\begin{aligned}F & =\left|\frac{\mathrm{Tr}\left[R_{\overrightarrow{n}}^{\dagger}(\theta_{T})R_{\overrightarrow{n}}(\theta_{t})\right]}{2}\right|{}^{2}\\
 & =\left|\mathrm{cos}(\frac{\triangle\theta}{2})\right|^{2}.
\end{aligned}
.\label{eq:9}
\end{equation}
Here $\triangle\theta=\theta_{T}-\theta_{t}$. Apparently, we can
get the following relation 
\begin{equation}
\triangle\theta=2\mathrm{arccos}(\sqrt{F}).\label{eq:10}
\end{equation}
Then we can obtain the required subsequent rotation angle $\triangle\theta$
via the current fidelity in the coaxial case. 

\begin{figure}[!htbp] 
\centering
\includegraphics[scale=0.28]{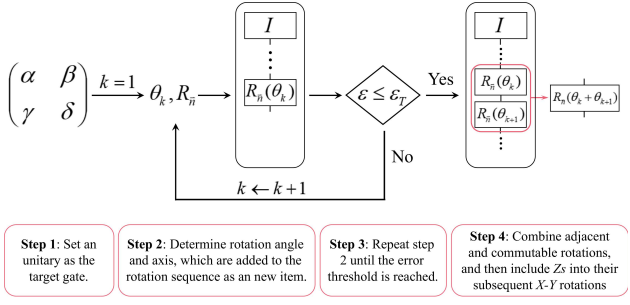}

\caption{(Color online) Workflow of using $SN$ algorithm for approximately compiling arbitrary
single-qubit gates.}\label{Fig.1}

\end{figure}

However, the presumed rotations around the axis $\overrightarrow{n}$
in Eq.~(\ref{eq:8}) may not be experimentally accessible
in the underlying platform. Our strategy for this problem  is
to substitute an experimentally permitted approximate axis for this theoretical one, where the 
approximation behaves with the best fidelity among rotations about
all possible axes with the same angle $\triangle\theta$. Then with the corresponding unitary
obtained, a new rotation angle and approximate axis can be performed
again in the same way, until the gate error $\epsilon=1-F$ is smaller
than a desired target value $\epsilon_{T}$. All of these approximate rotation operations eventually form a sequence in order. Then the adjacent and
commutable rotations in the sequence can be merged together to reduce the
redundant pulses. At last, the $Z$ rotations are absorbed into the subsequent
$X$-$Y$ operations by taking advantage of the virtual $Z$ technology. The workflow of this algorithm
is shown in Fig.~\ref{Fig.1}.  We point out that the appropriate rotation axes and angles can be determined by this algorithm, then arbitrary single-qubit unitary operator is compiled dynamically and automatically. We call it the self-navigation ($SN$) algorithm. 

\begin{figure}[!htbp] 
\centering	
\subfigure[ ]{\includegraphics[scale=0.21]{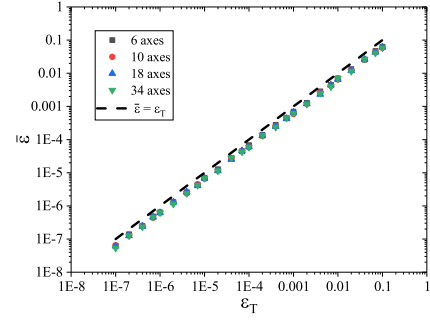}}
\subfigure[ ]{\includegraphics[scale=0.21]{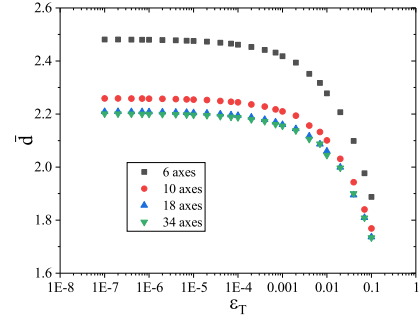}}
\subfigure[ ]{\includegraphics[scale=0.21]{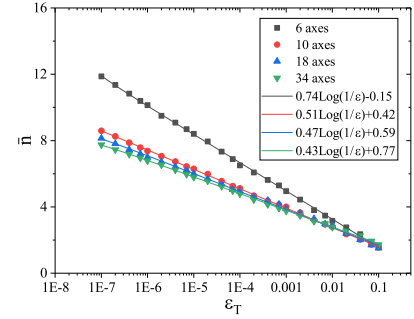}}
\subfigure[ ]{\includegraphics[scale=0.21]{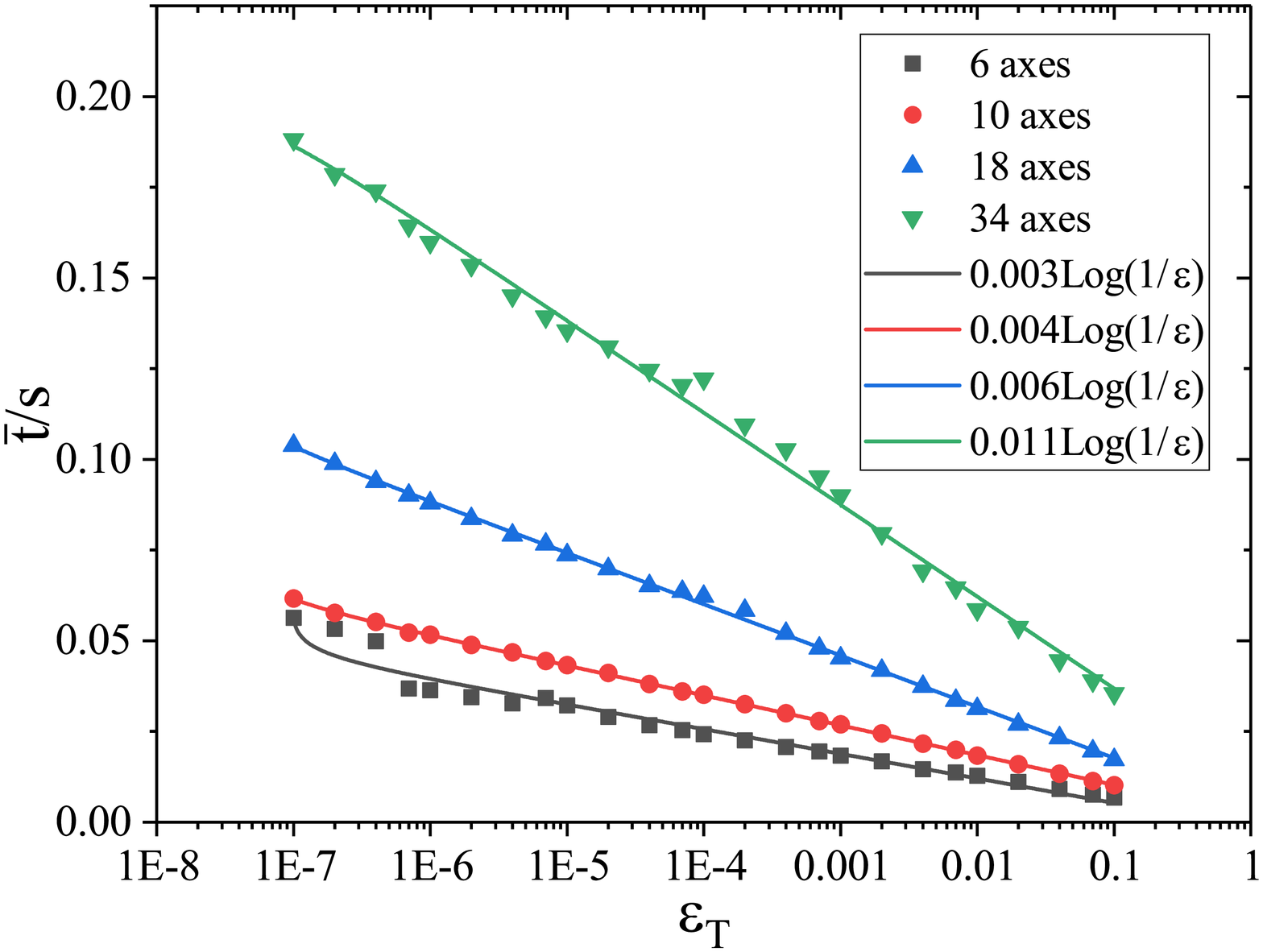}}

\caption{(Color online) The average actual accuracies $\bar{\epsilon}$ (a), averaged overall
rotation distances $\bar{d}$ (b), averaged numbers of pulses $\bar{n}$, and the averaged time $\bar{t}$ (d) versus the desired
target accuracy $\epsilon_{T}$ for different allowed rotation axes. The evaluation is achieved by the $SN$ algorithm. The auxiliary dashed line in (a) indicates
the threshold $\bar{\epsilon}=\epsilon_{T}$. The solid
curves in (c) and (d) are the fitting functions based on the corresponding
actual data.}\label{Fig.2}

\end{figure}

To evaluate our algorithm, as an example we explore its performance in the context
of superconducting transmon. We exemplarily take $\pm z$-axes
and several axes located in the $XY$-plane as the approximate rotation
axes. All these axes are accessible on the  platform. And they can be parameterized by modifying the phase offset $\phi$ in Eq. (\ref{eq:2}).
As we can take any rotation around the $\pm z$ and axes in $XY$-plane
as the approximate operations by use of Eqs.~(\ref{eq:1}) and (\ref{eq:2}),
we study the performance of the $SN$ algorithm with different number of
allowed rotation axes. Specifically, we take $N_{\mathrm{axes}}=6$, $10$, $18$ or $34$ as an example.
In each case, the $N_{\mathrm{axes}}-2$ axes which is distributed uniformly on the $XY$-plane and  the $\pm z$ axes  are embraced. The $N_{\mathrm{axes}}-2$ axes are characterized by the phase offset $\phi\in[0,2\pi)$ in Eq. (\ref{eq:2}). Then the rotation axes $\pm x$ and $\pm y$ will always exist.  It is easy to see that the algorithm can easily compile typical single-rotation gates around these axes, i.e.,
$X_{\theta}$, $Y_{\theta}$ and $Z_{\theta}$. The evaluation dataset
consisting of 128 single-qubit target gates is formed by two successive
rotations around the $x$- and $z$-axes with $\theta$ and $\varphi$
angles respectively. $\theta\in[0,\pi]$ and $\varphi\in[0,2\pi)$.
8 $\theta$ and 16 $\varphi$ angles are uniformly sampled from their
respective domains.

In Fig.~\ref{Fig.2} (a)-(d), we plot the average actual accuracies $\bar{\epsilon}$, overall rotation distances $\bar{d}$, numbers of pulses $\bar{n}$, and runtime $\bar{t}$ as functions of the desired target accuracy $\epsilon_{T}$ for different allowed axes. The evaluations are all obtained by our $SN$ algorithm.
The dashed line indicates the threshold $\bar{\epsilon}=\epsilon_{T}$ in Fig.~\ref{Fig.2} (a). Obviously,  Fig.~\ref{Fig.2} (a) shows that any given approximate accuracy can be achieved when the $SN$ algorithm
terminates for all cases.  From Fig. \ref{Fig.2}
(b) we can easily conclude that $\bar{d}$ varies with different
$N_{\mathrm{axes}}$  and tends to converge to certain values as $\epsilon_{T}$ decreases,
e.g., 2.2 for $N_{\mathrm{axes}}=18$. This value is significantly smaller
than the fixed $\pi$ obtained by the $U3$ gates \cite{IBM_U3,HUAWEI_U3,virtual_Z_gate}.
In addition, we find that the performance of the $SN$ algorithm is enhanced
with increasing $N_{\mathrm{axes}}$, and
this trend gradually weakens and disappears when $N_{\mathrm{axes}}>18$.
Thus it is enough to use 18 allowed axes in this compiling task, and more axes yield only negligible
returns. Fig. \ref{Fig.2} (c) clearly shows that $\bar{n}$ scales up with $\mathcal{O}[\mathrm{Log}(1/\epsilon)]$
as $\epsilon_{T}$ decreases. In addition, as expected the more allowed axes can
be performed, the less pulses is required. Notably, $N_{\mathrm{axes}}=18$ is sufficient for the
implementation.  The efficiency is also an important metric to evaluate an algorithm.
Fig. \ref{Fig.2} (d) shows that more axes will result in
longer average runtime $\bar{t}$. $\bar{t}$ roughly scales up with
$\mathcal{O}[\mathrm{Log}(1/\epsilon)]$ with very small prefactors (about $10^{-3}\sim10^{-2}$)
as $\epsilon_{T}$ decreases for all cases. In short, comparing
the polylogarithmic ($\mathcal{O}[\mathrm{Log}^{c}(1/\epsilon)],c\sim3$) overhead cost
with Solovay-Kitaev algorithm \cite{SK_algorithm}, our $SN$ algorithm
can approximate arbitrary single-qubit gate to any accuracy with
logarithmic cost both in terms of the required pulse number and
the total design time.

Given the limitations of quantum computing hardware presently accessible,
we simulate quantum computing on a classical computer and generate
the corresponding data. Our algorithms are implemented with PYTHON
3.7.9 and run on a computer with four-core 1.80 GHz CPU and 8 GB memory.
The source code and detailed data supporting this work can be found
in Ref.~\cite{code_GC_via_SN_algorithm}.

\section{Conclusions and discussions}

In practical quantum computing, the physical qubits will inevitalbly suffer from external noises, causing the population leakage and limiting
the coherence time. One strategy to protect the qubit is utilizing gentle pulses, whose spectrum components
on leakage are highly suppressed. Simultaneously, the
gate time is shortened by taking a short rotation path. The approximate quantum gate compiling is a promising approach to
get a better performance with contemporary NISQ device. 
In this work, we propose a $SN$ algorithm to approximately compile arbitrary single-qubit gate with
rotations all natively available in the experiments. The evaluation
results show that the overall rotation distance generated by our
$SN$ algorithm is significantly shorter than the
commonly used $U3$ gate. In addtion, any accuracy to
the target gate can be achieved by our $SN$ algorithm at the expense of the increasement of the pulse number and design time of the scheme. Combined with the virtual Z gate
technology, we find that both the two above values are
modest and only logarithmic in $1/\epsilon$ with very small prefactors,
denoting low overhead costs. We also show that, with 18 allowed
rotation axes, the $SN$ algorithm is ``just enough'' to balance the
gate performance and the control cost. We emphasize that
although our discussions focus mainly on the superconducting transmon
qubit, the techniques introduced here can be applicable
to a wide array of physical systems. 

We only consider the single qubit gate compiling in this paper. We expect our work could promote the further research of multi-qubit gates. For example, in error correction coding of taking the factoring a large number into its primes, a faster gate operation offered by the
shortened control path means smaller qubit overhead. This demanded to generate and purify the special ancilla states which are used to construct the Toffoli
gate. In addition, due to the highly sensitivity to the error rate
in physical qubits, the improved quantum control over the physical
qubit will further significantly reduce the size of the circuit.

\section*{Acknowlegment}

This work was supported by the Natural Science Foundation
of Shandong Province (Grant No. ZR2021LLZ004), and the Natural Science Foundation of China
(Grant No. 11475160). RHH would also like to thank Sheng-Bin Wang, Zhi-Min Wang, and
Guo-Long Cui for fruitful discussions.

\bibliography{References_library.bib}

\end{document}